\documentclass[final,3p,times,twocolumn]{elsarticle}

\usepackage{graphicx}
\usepackage{amssymb}

\usepackage{ulem}

\biboptions{sort&compress} 

\usepackage[hyphens]{url}

\usepackage{xcolor}

\usepackage{comment}
\usepackage{enumitem}
\usepackage{lineno}

\journal{Nuclear Instruments and Methods in Physics Research Section A}

\begin{document}

\begin{frontmatter}


\title{
Instrumenting a Lake as a Wide-Field  Gamma-ray Detector
}


 \author[MPIK]{Hazal Goksu}
 \author[MPIK]{Werner Hofmann}
 \author[MPIK]{Felix Werner}
 \author[MPIK]{Fabian Haist}
 \author[MPIK]{Jim Hinton}
 \address[MPIK]{Max-Planck-Institut für Kernphysik, Saupfercheckweg 1, Heidelberg 69117, Germany.}


\begin{abstract}

Ground-level particle detection has recently emerged as an extremely powerful approach to TeV-PeV gamma-ray astronomy. The most successful observatories of this type, HAWC and LHAASO, utilise water-Cherenkov based detector units, housed in tanks or buildings. Here we explore the possibility of deploying water-Cherenkov detector units directly in to a natural or artificial lake. Possible advantages include reduced cost and improved performance due to better shielding.  The lake concept has been developed as an option for the future Southern Wide-view Gamma-ray Observatory, and is now under consideration for a possible future extension of the observatory, beyond its recently selected land site. We present results from prototypes operated in a 
custom built facility, and concepts for full-scale array deployment and long-term operation. 

\end{abstract}


\end{frontmatter}

\nolinenumbers





\section{Introduction}

 
Ground-level particle detection, for the purposes of gamma-ray astronomy at TeV energies, requires both high fill-factor and large area ($10^{5}$\,m$^{2}$). As a consequence individual detector units must be heavily cost optimised in terms of cost per m$^{2}$ at fixed performance. Water Cherenkov detectors (WCDs) are very promising in this regard, providing conversion of shower photons and calorimetric response, utilising a rather low cost medium (pure water). However, in existing systems the containment of the water typically represents a very significant share of the total cost. For example in the HAWC Observatory in Mexico, large (7~m diameter) steel tanks were adopted to reduce per m$^{2}$ cost, but still dominate over the photosensor, electronics and water costs~\cite{Abeysekara_2017}. In LHAASO a different approach was adopted, providing very large fill factor using large water-filled buildings with optical separators, but also with significant associated cost and complexity, associated to keeping the water clean and managing humidity within the buildings, as well as dealing with leaks \cite{2022ChPhC..46c0001M}. 

The fundamental needs for a WCD are sufficiently transparent water (e.g. with absorption length significantly larger than detector size), blocking of external light with an attenuation of $10^{18}$ 
and mechanical support. An external volume of water, such as a lake, can potentially provide the required mechanical support at modest cost. The remaining requirement of a light and water tight bag (or bladder), that does not pollute the water within or impact the lake environment, may potentially be met at a fraction of the cost of a lined tank or building.

The idea of using a lake for astroparticle physics is not new: in particular for neutrino detection there is a long heritage and a major facility in Lake Baikal~\cite{1997APh.....7..263B}, but also air-shower detection has been discussed, for example acoustic detection of UHE cosmic rays in Lake Titicaca~\cite{1983ICRC...11..428K}. Only in the context of the SWGO project~\cite{Hinton_2021}, however, has a lake solution been evaluated for gamma-ray astronomy~\cite{Goksu:2021Tw,Goksu:2023Pk}~\footnote{UHE gamma-ray astronomy with muons, detectable in a wide range of detector types including neutrino detectors, including the one in Lake Baikal, has been proposed by Halzen et al~\cite{1995hep.ph....7362H}. However, the technique has not been demonstrated in practice, nor any dedicated facility proposed. The MACRO experiment later searched for an excess of muons of celestial origin, obtaining only upper limits~\cite{1993ApJ...412..301A}.}. Several large high altitude lakes exist in Peru and Bolivia, and three in the Sibinacocha region of Peru have been considered for SWGO~\cite{2022icrc.confE.689D, Abreu:2023WI}. Recently, the site of Pampa la Bola in Chile, a site without nearby lakes or rivers, has been selected for SWGO, hence the possibility of using a natural or artificial lake for the main array is no longer possible. The lake deployment of detector units is now considered for a possible future multi km$^2$ extension.

An artificial pond or lake, created for the express purpose of gamma-ray astronomy, would of course add significant cost w.r.t. to a natural lake, but may still be cost-effective in comparison to individual tanks, depending on the local environment and civil engineering work required~\footnote{The concept of using closed ponds was previously proposed with the Gamma-Ray And Neutrino Detector (GRANDE) experiment, envisioned as a water Cherenkov detector that serves as an extensive shower array and neutrino detector, utilizing flooded quarries~\cite{1990NuPhS..14..125S}. The Milagro experiment was later constructed using a large pond with a light-tight cover, becoming the first large-scale water Cherenkov detector observatory,~\cite{2005ICRC...10..227S}. 
}. In a region in which evaporation from an open pond can be readily offset with natural/local water sources, the solution presents several potential advantages with respect to a closed building in terms of cost and complexity. An artificial water volume may of course also present significant advantages over a natural one - in terms of ease of access and risks from waves for example. 

In 2020 we constructed a test facility at the Max-Planck-Institute for Nuclear Physics (MPIK) in Heidelberg, Germany to evaluate the lake concept for future use in a gamma-ray observatory. Here we report on our evaluation of the concept using this facility, on simulations of performance differences for a lake-based solution and on considerations for implementation of an array based on this concept.

\section{Detector Unit Design Concept}  \label{sec: design}

The overall concept for a gamma-ray observatory assumed here is of independent light-tight water Cherenkov Detector units (WCDs). Signals from photosensors in each WCD are used for air-shower detection by the array as a whole. Following~\cite{2023NIMPA105068138K}, we consider a double-layer WCD(DLWCD) with a lower chamber focused on muon tagging. White/reflective walls are desirable at least for the lower chamber and likely for some surfaces of the upper unit as this increases detection efficiency. 

%


For a lake-based array, the WCDs float in the lake and are coupled together into clusters or islands for positional stability. These clusters would have a modest size ($\sim7$ detector units), but would be flexible. The arrays could be rearranged by moving the detector units and anchors with a boat. In case of a need for a high fill factor, long clusters of two detector units would be reasonable. Below we consider the details of the design and its advantages, before considering specific requirements on materials and the implications of wave motion.
A hierarchical solution in terms of data and power distribution is envisaged -- with a {\it DAQ field node} distributing power to, and collecting data from, $\sim$100~WCDs. 

\begin{figure*}
    \centering
    \includegraphics[width=\linewidth]{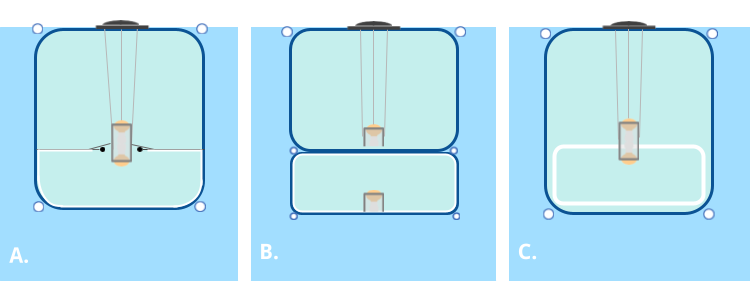}
    \caption{Three design options, from left to right: A) internal divider, B) independent volumes, C) the matryoshka, or nested solution.}
    \label{fig:bladdersabc}
\end{figure*}

\begin{figure}
    \centering
   \includegraphics[width=0.95\linewidth]{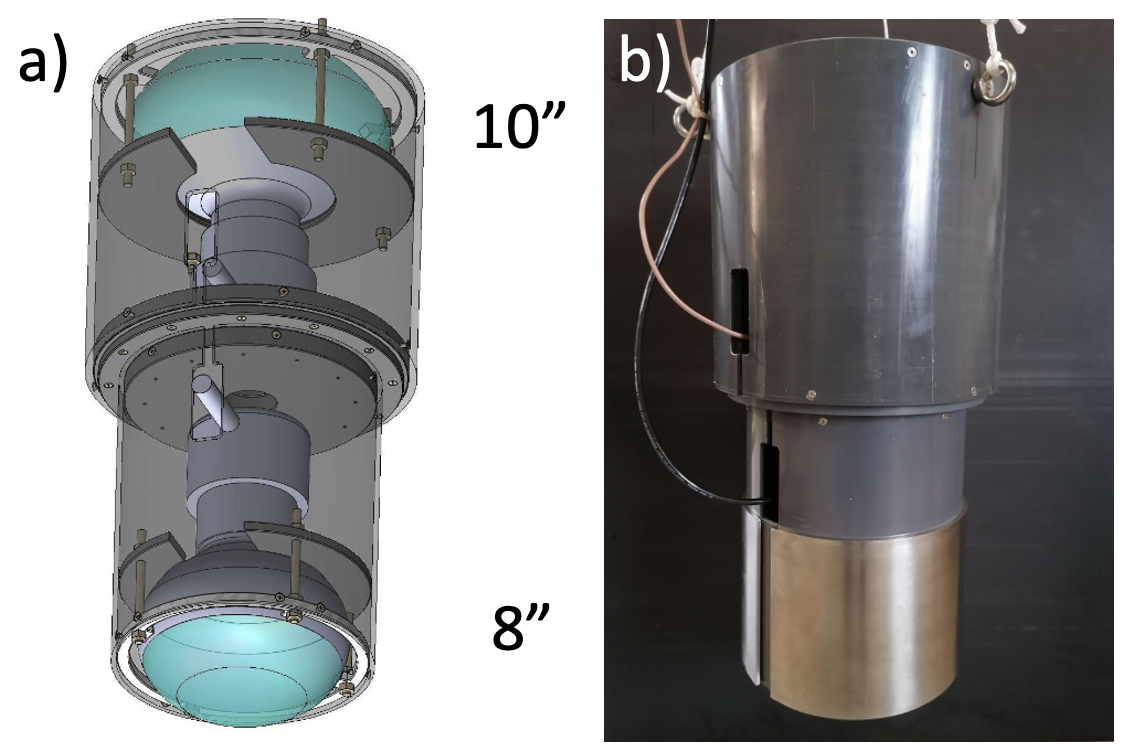}
    \caption{Double photo-multiplier assembly for installation in a WCD: a) mechanical design, b) prototype with a 10-inch Hamamatsu R7081 PMT facing up and an 8-inch R5912 PMT facing down. The mass of the unit is 21\,kg and the net-negative buoyancy about 2\,kg.}
    \label{fig:doublePMT}
\end{figure}

Figure~\ref{fig:bladdersabc} shows three possible solutions for the implementation of a two-layer WCD in a lake. In all cases, a photosensor or pair of sensors -- as illustrated in Fig.~\ref{fig:doublePMT}  -- is hung, as a unit of net negative buoyancy, from the access hatch of the WCD. The cases are:
\begin{enumerate}[label=(\Alph*)]
     \item  A single outer bladder with an internal divider. The photosensor module is lowered in to a gap between the two layers.
     \item  Two entirely separated bladders are deployed together, with photosensor pre-integrated in to the lower bladder. The photosensor in the lower bladder may be attached so that it faces upward or downward. 
      \item  The Matryoshka, or nested solution, where a reflective inner bag is deployed inside the outer bladder.
\end{enumerate}

Advantages of A) include ease of deployment,
but it may not be trivial to manufacture with the required reflectivity of all surfaces.
B) allows the size and spacing of the lower level units to be decoupled/separately optimised with respect to the upper units, but may be non-trivial to deploy. C) has advantages in terms of manufacture and deployment, provided a reliable mechanism can be implemented for unfolding the inner bag. Section~\ref{sec:exp} describes a prototype based on option C.

\begin{figure}
    \centering
    \includegraphics[width=0.7\linewidth]{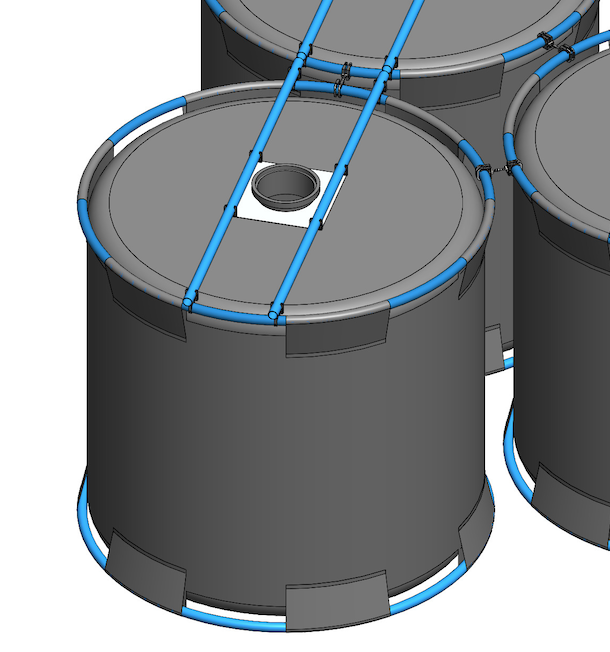}
    \caption{Bladder with floater and stretcher ring, and the optional connection between hatch and floater ring.}
    \label{fig:bladdersdetail}
\end{figure}

All prototype designs have in common the use of an air-filled or foam-filled PE ring pipe attached to the top of the bladder, serving to keep the bladder afloat (``floater ring'') and the use of a second ballast-loaded PE ring at the bottom  (``stretcher ring'') serving to define and stabilize the bladder shape. The rings are made of commercial low-cost black PE pipe, and attached to the bladders using sleeves made from bladder material and welded to the bladder body (see Fig.~\ref{fig:bladdersdetail}). For large-scale production, use of custom blow-molded floater elements is conceivable, as used e.g. for floating solar arrays.

The hatch is equipped with floaters to prevent permanent immersion of the hatch seal in the lake; various options were investigated, with floaters inside the bladder, or around a hatch neck. Optionally, the hatch can be connected to the floater ring via diagonal PE pipes, as shown in Fig. \ref{fig:bladdersdetail}.

\subsection{Advantages of the Lake Approach}

In general, the lake concept offers the promise of extended flexibility, with an array that can be tuned for scientific needs, combined with reduced costs. Specific advantages are:

\begin{itemize}
    \item The material cost of a lake-based unit is reduced due to the absence of a water container or tank, the additional needs in terms of flotation are modest in comparison. However, additional costs may be introduced due to the deployment requirements that may arise. The costs per detector unit are expected to be dominated by the bladder and its hatch, with additional costs from floaters, bladder interconnects and a position monitoring system; and finally the deployment and filling of detector units on site.  In Section \ref{sec:deployment} we suggest a few methods of deployment expected to have a limited cost. 

    \item The lake approach ensures a flexible array configuration that can be modified according to the changing needs of the observatory over time; e.g. a higher energy or a lower energy focused configuration, as the individual WCDs can be readily re-positioned without removing water or sensors, unlike the case of a tank-based array.

    \item As there are no tanks or enclosures to account for, the individual detector units can be optimized entirely based on physics requirements, as opposed to taking into account other factors such as the size of commercially available tanks. 

\item The presence of water around each detector unit will suppress sideways entry of electromagnetic particle, likely improving the quality of muon tagging using the lower chamber. 

\item Logistics are simplified: the amount of material to ship or transport to the site would be relatively small compared to tank solutions. 
\end{itemize}

\subsection{Challenges} \label{subsec:challenges}
  
As this is a new technology, no previous experience exists for large-scale deployment of lake detectors. This means every complication that could occur needs to be evaluated, by working with realistic prototypes and mini-arrays. Specific challenges include:

\begin{itemize}
    \item The mechanical stability of detector units under water motion needs to be ensured. In the presence of wave loads, the bladders are expected to move and they should be able to withstand the forces such movements would bring. Small movements of bladders against one another over time could also potentially damage them over long periods, as can fatigue due to continuous motion. Single and small array bladder units should be tested in a controlled environment against wave motion.

\item The absence of an outer container increases the requirements on the bladder, in terms of durability, tolerance to long-term UV exposure and light-tightness. Bladders must not contaminate the water they are deployed in, but this is essentially guaranteed by the requirement that exists for all WCD bladders, to preserve the water quality inside.

\item Since the units will be floating in water and will be subjected to disturbances, the geometry calibration is more complex than the static tank case. Strategies for monitoring the position of the bladders, and the spatial orientation and position of the photo-multiplier tubes (PMTs) inside each bladder need to be developed. These could be data-driven position calibration methods that are augmented by specific instrumentation.

\item Challenges exist regarding the identification of a suitable site. If a natural lake is preferred, this greatly reduces the choice of sites. Furthermore, environmental and cultural concerns beyond water contamination may arise. Especially in the case of a natural lake, a detailed assessment of the environmental impact on local wildlife and vegetation would need to be carried out with local experts. High altitude natural lakes might house fragile ecosystems, including unique species of flora and fauna that may not be found elsewhere. It is crucial to protect these ecosystems by ensuring that the installation of a WCD array does not disrupt the natural habitat. For instance, measures should be taken to prevent any alterations to the water quality or the physical environment that could harm local wildlife or plant life. If an artificial lake is preferred, this brings about extra engineering efforts and costs. Restrictions on array geometry are also expected, since in addition to the dense array, the sparse array would need to be arranged.

\end{itemize}

\subsection{Material Requirements} \label{sec:materials}

The individual detector units need to be made of materials that can remain light tight and water tight over a period of several years. In the lake deployment case the bladders would not have additional structural support from tanks, hence, the bladder material is crucial for the stability and performance of the detector units. This material needs to meet several requirements including mechanical strength and durability, light tightness, reflectivity of the inner liner, and lack of contamination of the purified water inside (and the lake water outside) the bladder. Polyethylene liners made up of several layers for different purposes appear as the best candidates, similar to what is done for the Auger Observatory \cite{2008NIMPA.586..409P}, for HAWC Observatory \cite{2023NIMPA105268253A} and for the muon detectors of LHAASO \cite{2022ChPhC..46c0001M}. To achieve high reflectivity, a Tyvek-layer can be laminated to the liner \cite{2008NIMPA.586..409P}.

The strength and durability of the bladder materials are essential to keep the detectors intact. The bladder materials should stay within their elastic range under the influence of wave motion, both in the short term and over many years. Flexibility and elasticity are important for the bladders to stay intact and tear-free. Tests to ensure there would be no tearing or contamination of water should be conducted using samples of candidate materials. The outer layers of the materials should be able to withstand the UV radiation to which the top of the bladder may be exposed. Any gradual contamination of the water inside the detector units can be monitored using muon signals (see~\cite{WANG2020163416}). 

Light tightness of the bladders is crucial for the detection of Cherenkov light from the incoming particles. The materials and bladders that are durable and flexible should be tested for light tightness at every stage. It should be ensured that the packaging and folding of bladders does not diminish their light tightness. To this end, ensuring that the bladders are made with strong seams is very important.

The lower chamber requires a liner with high diffuse reflectivity for blue/near-UV light. A reflective lining for the lower chamber maximizes signal efficiency depending on its reflectivity. However, some liners tend to deteriorate the quality of the water they are in contact with, causing lower absorption lengths. This diminishes the signal obtained from the WCDs, in particular in the lower chamber where multiple reflections lead to long path lengths. Through water quality tests, we have observed that materials like polyvinyl chloride (PVC) can quickly degrade water quality, whereas low-density polyethylene (LDPE)—used in the detector units of the Auger Observatory, HAWC observatory and LHAASO— and high-density polyethylene (the material Tyvek is made from) preserve a high absorption length. 

The liner being reflective results in a wider spread in photon arrival times, leading to a worse time resolution (see~\cite{2023NIMPA105068138K}); however angular reconstruction for the WCD array is anticipated to be based only on the timing information from the upper chambers.

During our prototyping studies we have used different setups to check these requirements. To assist with water quality checks, we have been operating a screening facility for water quality deterioration since May 2021, housing different material samples inside buckets filled with purified water and using a Sea-Bird C-Star transmissometer at the wavelength of 410\,nm to monitor water degradation. Furthermore, we measure the reflectivity of candidate liners with an integrating sphere coupled to a monochromator. We have also tested prototype bladder materials for light tightness using PMTs, a lamp with 35 W power output (Around $10^{18}$ photons/m$^2$ output) and natural sunlight, where we found the materials to meet our light attenuation requirement. Although not suitable for mass use, these setups served as guides during prototyping. In the case of mass production, dedicated and automated setups  and procedures for quality control of the bladders are needed.

\subsection{Water Motion}


An array of water Cherenkov detector units deployed in a natural lake would be subjected to the water motion, potentially causing deformations in bladder shape and positions of the light sensors. As mentioned in section \ref{subsec:challenges}, concerns include the continuous mechanical load produced by waves, fatigue of the bladder materials, the ability of bladder materials to withstand the maximum deformations resulting from large waves, and the effect of movement of light sensors inside the bladders on shower reconstruction. 

\begin{figure}[h!]
    \centering
    \includegraphics[width=\linewidth]{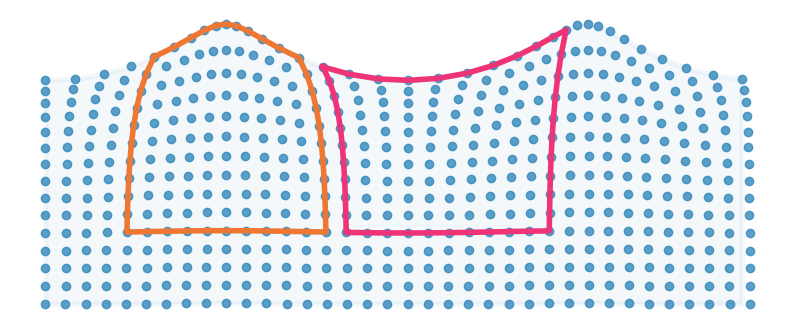}
    \caption{Sketch of trochoidal (idealised case for in-compressible fluid) wave motion, illustrating deformation modes for lake-deployed bladders. The largest forces exist close to the surface and for long wavelengths and large amplitudes.}
    \label{fig:waves}
\end{figure}

Behavior of water waves is explained to a linear approximation via the Airy Wave theory, which describes the free surface elevation of water waves as a sine wave. \cite{waves_coastal, coastal_engineering, oceanography}. Under the free surface, the fluid particles follow closed circular orbits with radii that decrease with distance from the water surface, as $R = a e^{2\pi z / \lambda}$, where $a$ is the wave amplitude, $\lambda$ is wavelength and $z$ is the distance from water surface (with $z < 0$ below the surface). This implies that the influence of wave-induced particle motion diminishes with depth, suggesting that positioning detectors at a depth of approximately 2 meters or more can mitigate wave effects. However, deeper placement of detectors leads to increased absorption of low-energy electrons and, at greater depths, substantial attenuation of electromagnetic energy, as illustrated in Figure~\ref{fig:overburden_gammas}


\par Although the reality is more complicated, Airy theory is sufficient to illustrate that waves would cause stretching and deformations in bladder shape, as shown in Fig. \ref{fig:waves}.  Given the large forces associated with wave motion, the bladder shapes will basically follow the water motion; only across many bladders, towards the inner region of a dense array, some wave damping will occur. Tensile strength and in particular elasticity of bladder materials must be such that bladders do not rupture under such deformations. PE foils are materials that can stretch many times their length without breaking, hence they are highly suitable as bladder materials. 


To further reduce water pressure build-up inside the bladder, bladders should not be filled to their maximum capacity; tests with smaller-scale prototypes point towards an optimum fill level of about 90\%, leaving a bit of slack in the bladder walls.


\par 
Wave heights (commonly defined as the vertical distance between the crest (peak) and the trough of a wave) and periods of wind-excited waves in lakes depend on wind speed $U$ and the size of the lake, which determines fetch $F$ (the distance over which wind interacts with the water surface). The expected wave heights and periods can be calculated using parameterised wave spectra under some assumptions, such as JONSWAP and Pierson-Moskowitz spectrum \cite{waves_coastal}.

\par Prior to deployment in any natural lake, depth surveys and an analysis of the weather conditions are necessary. In general, smaller lakes are favored due to the impact of fetch. For instance, lake Cocha Uma in Peru is at an altitude of 4800\,m and has a fetch around 0.8\,km. Assuming peak wind speeds of around 8\,m/s, we find using the formulas from \cite{master_norway} wave heights around 0.13\,m , wave length $\lambda=2$\,m and a period of around 1\,s. For larger lakes, approaching a fetch around 20 km, wave height could reach several meters, implying greater forces and bladder deformation.



The impact of waves can be minimised by careful array placement within the chosen lake - minimising the fetch in the prevalent wind direction - but assuming an unavoidable fetch of order 1~km and wind speeds up to 50~km/h, waves of wavelength up to 10 m and amplitude of up to 1~m must be survived. Bladder deformations for such waves should still be within the elastic limit of PE foil,
but tests with indoor wave pools, in large lakes, or at sea, are desirable to verify the survival of bladders. We note that the coupling mechanism between bladders and the anchoring scheme must be designed to avoid additional forces.




\section{Deployment and Operation of Detector Units} \label{sec:deployment}

Deployment of complete WCDs and operation of a lake-based facility long term represent significant challenges. Clearly schemes with highly reliable WCDs are favoured such that regular access to units is not required, none-the-less three aspects of access to/interaction with the array need to be considered:

\begin{enumerate}
\item {\bf Deployment}. Filling with water, deployment of photosensors, moving in to position, fixing in place.
\item {\bf Node maintenance}. Regular access for corrective or routine maintenance at floating DAQ field nodes will be required.
\item {\bf Access to individual WCDs} as part of a campaign of corrective maintenance that might be required on (for example) an annual basis.
\end{enumerate}

These aspects are addressed in subsections \ref{sec:depl}, \ref{sec:maint} and \ref{sec:ops} below. The operational aspects of position monitoring and calibration are included in Section~\ref{sec:ops}.

\subsection{Deployment}
\label{sec:depl}

Two methods of WCD deployment have been considered for the Lake concept: a) firstly filling with water and photosensor deployment at a pier, followed by moving the completed detector into place; b) in-situ filling from a boat with a reservoir of purified water, and a small crane to ease photosensor deployment. The pier based method is considered as the baseline. A water purification and storage facility is required at the shore, and a small crane on the pier. Bladders are prepared as a `flat-pack' with floater and stretcher rings tied together, for ease of lifting. The connection between the rings is then released once deployment into the water commences. Filling occurs with lids closed (using a custom lid for filling with hose connection). Once a fixed amount of pure water has been injected, the lid is removed and a photosensor or double-photosensors unit is lowered in to the filled bladder and secured at the neck. Signal and/or power cables are transported (coiled up) with the bladder, which is slowly towed to the final location. This deployment scheme was used and validated for prototype bladders deployed in the test tank at MPIK (7~m height, 10~m diameter). Once deployed, units are anchored to each other by connecting their floater rings using suitable flexible brackets. Anchoring to the lake bed is done for contiguous islands or rows of WCDs, with water ways between them for boat access. Floating DAQ field nodes collecting the signals from, and distributing power to, $\sim$100 WCDs are associated to each island, and could be deployed first to allow cabling to be done immediately.

\subsection{Maintenance}
\label{sec:maint}

Placement of field nodes at the edge of an island or row of WCDs ensures that they can be readily accessed by boat. Node units (occupying typically $\sim$1~m$^{2}$) can be placed on much larger floating platforms, for example occupying a full grid space for a WCD at modest loss of fill factor, to ensure that technicians can readily replace items in the nodes as needed. A heat exchanger coupling the node to the lake water through the platform can reduce the cost and complexity for temperature stabilisation of the node. In an artificial lake solution, nodes would likely be placed on land, at the edge of a pond. 

WCD access could be needed to replace a damaged/leaking bladder or faulty photosensors, or in-WCD electronics (although these elements should be designed for a MTBF well beyond the anticipated lifetime of the experiment). In the case of a solution based on long (full linear dimension of the dense part of the array) and narrow rows (e.g. four WCD units), only one WCD need be temporarily removed from the connection grid to access any other for bladder replacement (see Section \ref{sec:depl}). For photosensor access a temporary or permanent walkway is needed to reach and work at the relevant hatch. A customised blow-molded plastic floater solution is likely most cost effective and convenient at large-scale.

\subsection{Operation}
\label{sec:ops}

An array based on air-shower particle detection at ground level, with no moving parts, is highly suited to remote operation and monitoring. Very little is required in terms of operator intervention. One aspect that differs for a lake-based solution is position monitoring of WCDs/photosensors. For an artificial lake WCDs can be close-packed and no monitoring is needed. In a natural lake wave-motions and drift of clusters relative to anchor points will need to be monitored. Two concepts for this are considered, and combined if needed, depending on the results of field-tests:

\begin{itemize}
    \item To monitor their absolute positions and orientations, clusters could be fitted with multiple tracking markers, each of which are regularly scanned by robotic theodolites in a round-robin fashion. A drawback of this concept is the relatively low update rate and sensitivity to environmental conditions such as fog.
    \item Each bladder or cluster of bladders could be equipped with a GNSS sensor (rover) with integrated inertial navigation system (INS) for dead reckoning. The measurements of one or more static reference stations at the shore can be used to calculate correction data to achieve cm-level accuracy for each rover (differential GNSS). The INS data can then be used to provide orientation measurements and to interpolate between GNSS measurements to achieve update rates of several Hz.
\end{itemize}

In a data-driven approach, residuals of shower wave-front fits are used to derive a time correction and a correction of the effective position of a WCD units (by studying how fit residuals vary with shower direction). These corrections can than be referenced to the absolute position measurements discussed above. The high shower detection rate of dense WCD arrays allows determining sufficiently precise corrections on sub-second time scales. Given the typical achievable angular resolution in ground particle gamma-ray astronomy, of order 0.1 degrees, and considering shower baselines of 100-200 meters, a spatial resolution of a few tens of centimeters is necessary.






\section{Simulation of WCDs in a Lake} \label{sec:simulated}

The performance of a single unit detector immersed in water is in principle similar to a single unit detector on land, housed inside a tank. A detailed investigation of double-layer WCDs (DLWCDs) as individual detectors and as arrays is reported in \citep{2023NIMPA105068138K}, considering different chamber dimensions and liners. In this section, we evaluate only properties that would differ for DLWCDs inside a body of water; namely overburden and shielding. We use a simulation framework based on GEANT4 \cite{GEANT4_2002}, called HAWCSim, as used in \citep{2023NIMPA105068138K}. 
The WCD used in these simulations has a diameter of 3.6\,m, and upper chamber to lower chamber height ratio of $2.5:0.5$\,m, most similar to solution A) from section \ref{sec: design}. Two identical 8" R5912 photosensors are simulated and a water absorption length of 18 m at 410 nm is assumed, with the wavelength dependence as given by Segelstein \cite{Segelstein81}. The reflectivity of black polyethylene which is the inner liner for the upper chamber is assumed to be 0.1 while the reflectivity of Tyvek 1082D as the inner liner for the lower chamber is assumed to be 0.92, at 410\,nm, as measured by our MPIK setups. 

\subsection{Overburden}

\begin{figure}
    \centering
    \includegraphics[width=\linewidth]{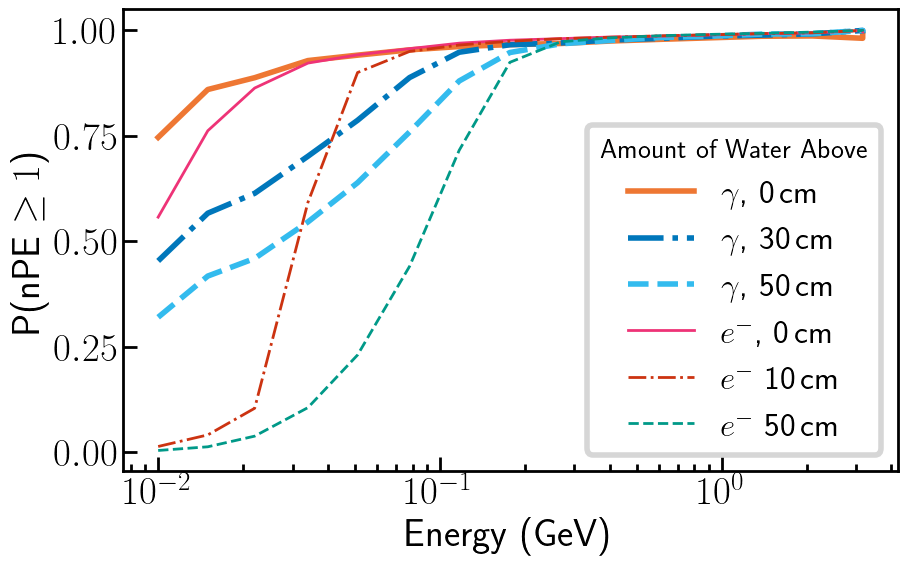}
    \caption{Impact of water depth above a WCD on the probability of detection of one or more photo-electrons as a function of particle energy.}
    \label{fig:overburden_gammas}
\end{figure}

 In the lake configuration, the top of the detector may be partially immersed in water, creating an overburden. Depending on the exact geometry and attachment of the flotation devices, detectors as described in Section \ref{sec: design} might have a few 10 cm of water above the sensitive volume. To test for this factor, a single DLWCD immersed in a body of water was simulated, with 0 to 50\,cm water overburden. The response to vertical gamma and electrons was simulated, for energies ranging from 10 MeV up to a few GeV. Fig.~\ref{fig:overburden_gammas} shows the detection probability versus energy of the incoming particle. In the case of low energy electrons, the detection probability drops sharply with overburden, while for gammas the impact of overburden is much reduced, as expected. Around GeV energies, overburden does not have a significant effect for gammas or electrons. Secondary particles with energies below 500 MeV have a decreasing probability of detection with increasing water depth, an effect similar to constructing the detectors at a lower altitude site, with each 10\,cm overburden of water corresponding to placing the detectors approximately 10\,m lower in altitude, slightly reducing the trigger efficiency. Given that most shower particles at detector level are gammas, and that shower particles below a few tens of MeV carry very little useful information, an overburden of 10\,cm is certainly tolerable.


\subsection{Shielding}

In the double chamber design, as described in Section \ref{sec: design} and in \cite{2023NIMPA105068138K}, the lower chamber serves to tag incoming muons, however electrons may still be able to enter this chamber from the sides, bypassing the upper chamber entirely. If the DLWCD is immersed in water, this affect would be avoided, providing shielding. We test for this affect by injecting gamma-ray initiated showers to a small (19 units) hexagonal array of DLWCDs with a minimum separation of 20~cm. CORSIKA showers with energies from 100\,GeV to 5\,TeV are directed to the center of the array, with different angles (at zenith, and at zenith angle $\theta=30^{\circ}$, azimuth $\phi=60^{\circ}$). 
\begin{figure}
    \centering
    \includegraphics[width=\linewidth]{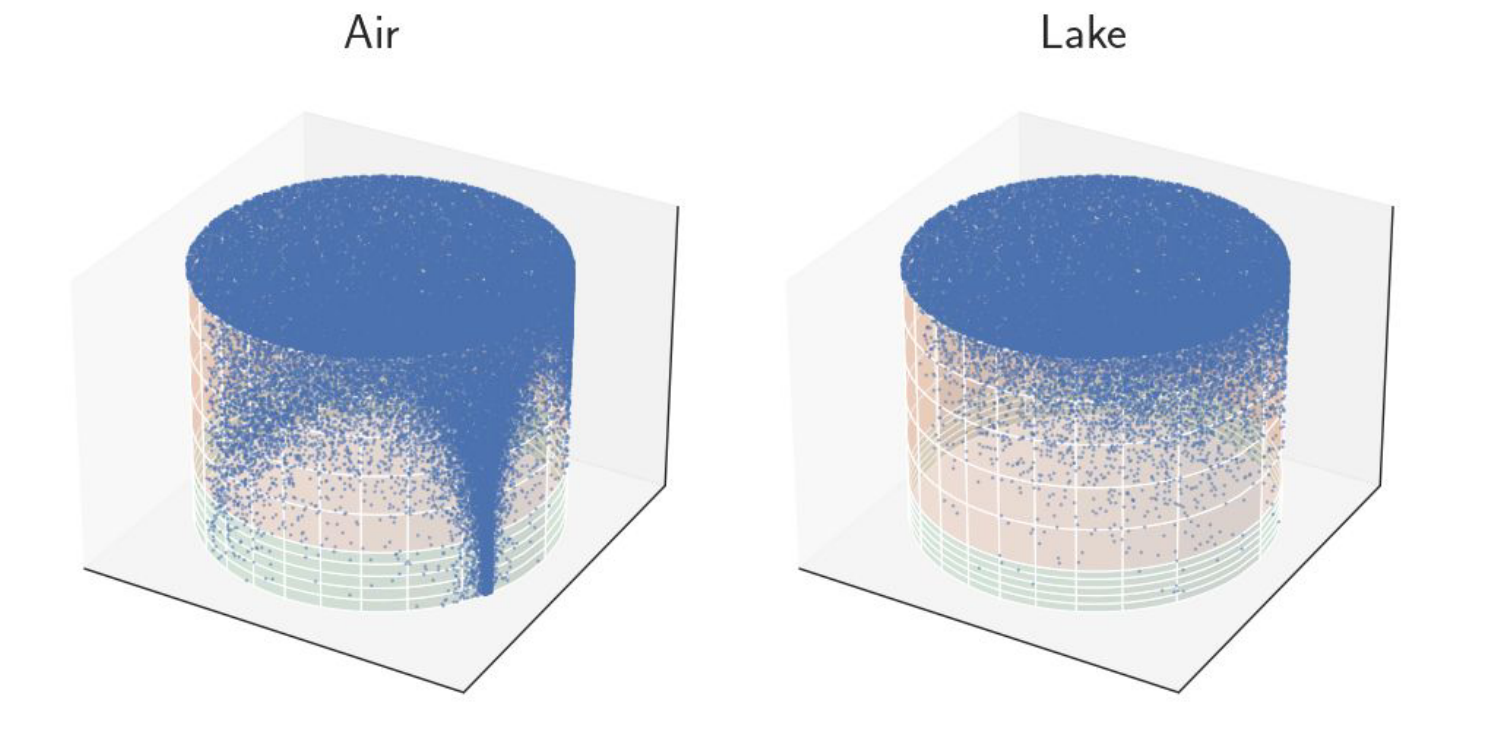}
    \includegraphics[width=\linewidth]{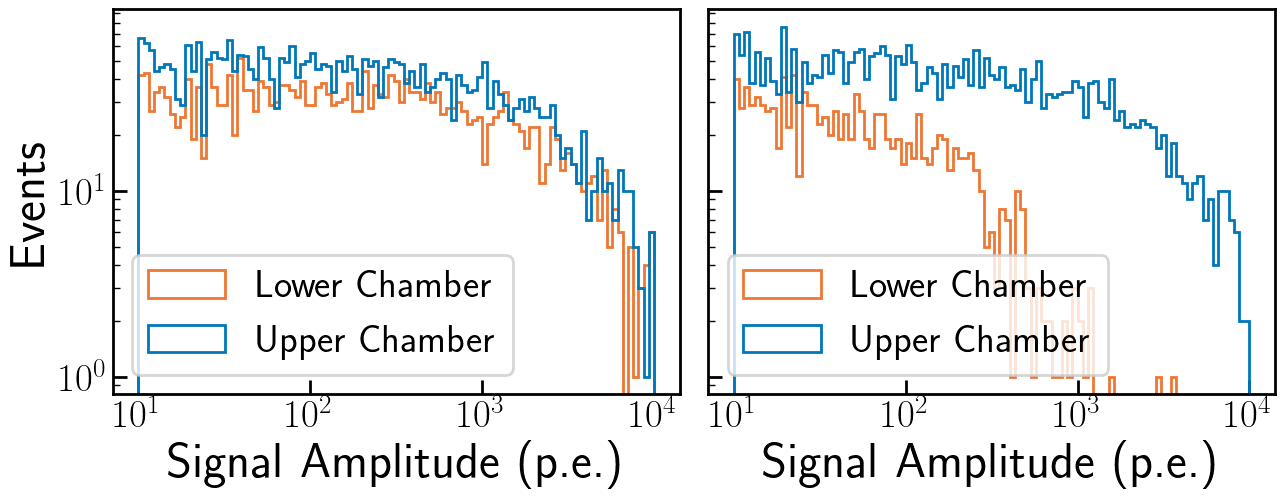}
    \caption{Effect of shielding for the central WCD in an array of WCDs in air \textit{(left)} and in water \textit{(right)}, for gamma-ray showers incident from $30^{\circ}$ zenith angle, directed to the central tank of the array. Top: locations of shower particles entering the WCD. In air, the lower chamber of the WCD is largely shielded by surrounding WCDs, except for particles penetrating in the gap between WCDs.
    Bottom: Signal distribution (in photoelectrons) in the upper and lower chamber, for the two cases. 
    }
    \label{fig:shielding}
\end{figure}

When showers are incident at non-zero zenith angle, the presence of water around the DLWCDs makes a difference. Fig.~\ref{fig:shielding} compares signals in the central DLWCD in the middle of the dense array in air with a similar DLWCD in water, for the case of $30^{\circ}$ inclined gamma-ray showers.  For the air case one sees the shadowing by the neighboring WCDs, shielding the central tank, but with particles penetrating in the gap between WCDs. When the unit detectors are immersed in water, sideways entry of particles into the lower chamber is much reduced. The effect is clearly visible in the amplitude distributions for the upper and lower chambers (bottom part of Fig.~\ref{fig:shielding}):
in air, the frequencies of large signals are similar for the upper and lower chambers, whereas in water, large signals are quite suppressed.
For gamma showers from zenith (not shown), the sideways entry into the lower chamber is quite small.
 The influence of this shielding on gamma/hadron separation and signal quality remains to be evaluated with full array simulations.



\section{Experimental Verification}
\label{sec:exp}
\begin{figure*}[ht]
    \centering
    \includegraphics[width=\linewidth,trim={0 0 0 1.5cm},clip]{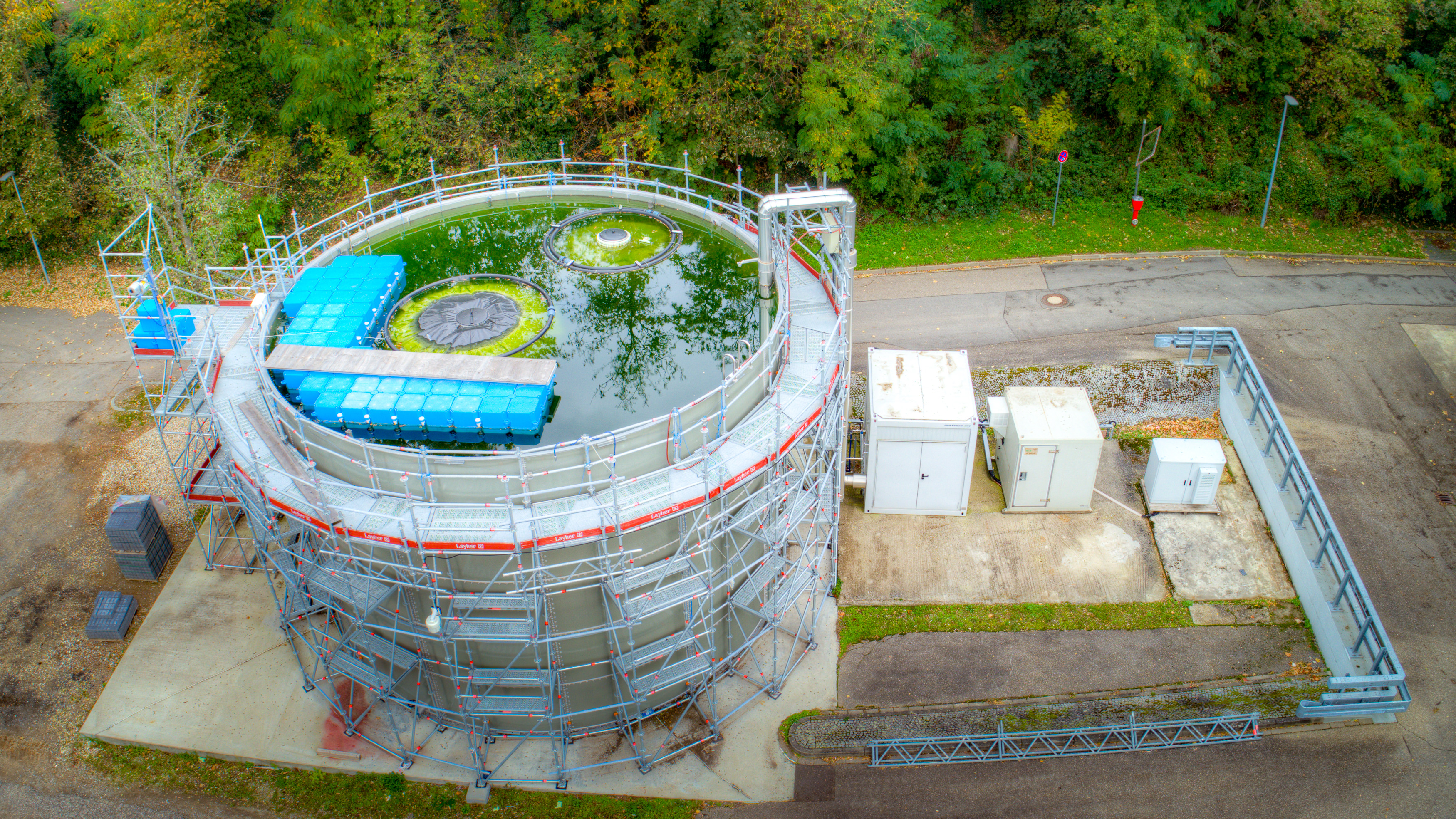}
    \caption{The test facility at MPIK: A 10~m diameter, 7~m height lake simulation tank, here with two bladders deployed. Next to the tank the water filtration system, and a small container for electronics and DAQ.}
    \label{fig:GWS}
\end{figure*}

A test pool of 7~m depth and 10~m diameter was constructed at MPIK in 2020, to facilitate the evaluation of lake deployment for WCDs (see Fig.~\ref{fig:GWS}). A filtration system provides filtered tap water for filling the bladders and an electronics cabinet houses a prototype SWGO electronics chain. This prototype electronics chain is a node based design with passive detector units, including the PMTs with a passive base, a coaxial cable, a data acquisition (DAQ) node~\cite{Werner:2021d/}.

In this test pool, a range of single-cell bladders made from different materials were deployed and tested, and  many of the mechanical concepts, such as the floater and stretcher rings, were explored and verified. A double-chamber bladder with internal divider (Scheme A of Fig.~\ref{fig:bladdersabc}) is under development at Aquamate\footnote{www.aquamatetanks.com}, but not yet delivered to MPIK. 
Instead, a prototype double-chamber bladder using an outer Polyethylene cell provided by Aquamate and a nested lower chamber custom-made at MPIK from Tyvek (Scheme C of Fig. \ref{fig:bladdersabc}) was used to perform mechanical tests and compare measurements in a test pool with simulated data.

\subsection{The Double-layer Water Cherenkov Detector Unit}
After testing several single-chamber prototypes to verify mechanics and deployment logistics, and to test the electronics chain, the double-chamber prototype based on Option C (Fig.~\ref{fig:bladdersabc}) was deployed inside the test pool. Although Option~A is likely more efficient for production in mass quantities, it also requires sophisticated industrial procedures for assembly and welding. Option~C on the other hand could be realized more quickly, using a normal bladder and an inner, Tyvek-only bladder that needs to be neither light-tight or watertight, and could easily be produced at MPIK. 

\begin{figure}[ht]
    \centering
    \includegraphics[width=\linewidth]{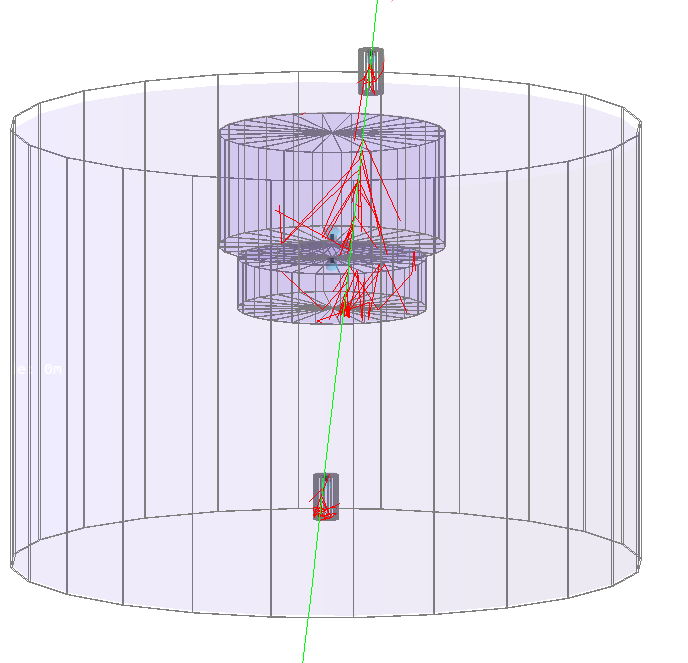}
    \caption{Visualisation of a GEANT4 simulation of the test facility in Heidelberg, where the lower chamber that is immersed into the larger bladder is depicted as a separate bladder with a smaller radius}
    \label{fig:simsetup}
\end{figure}
\begin{figure*} 
  \centering
      \begin{tabular}[b]{c}
    \includegraphics[trim={0 0 0 0 cm},clip, width=.38\linewidth]{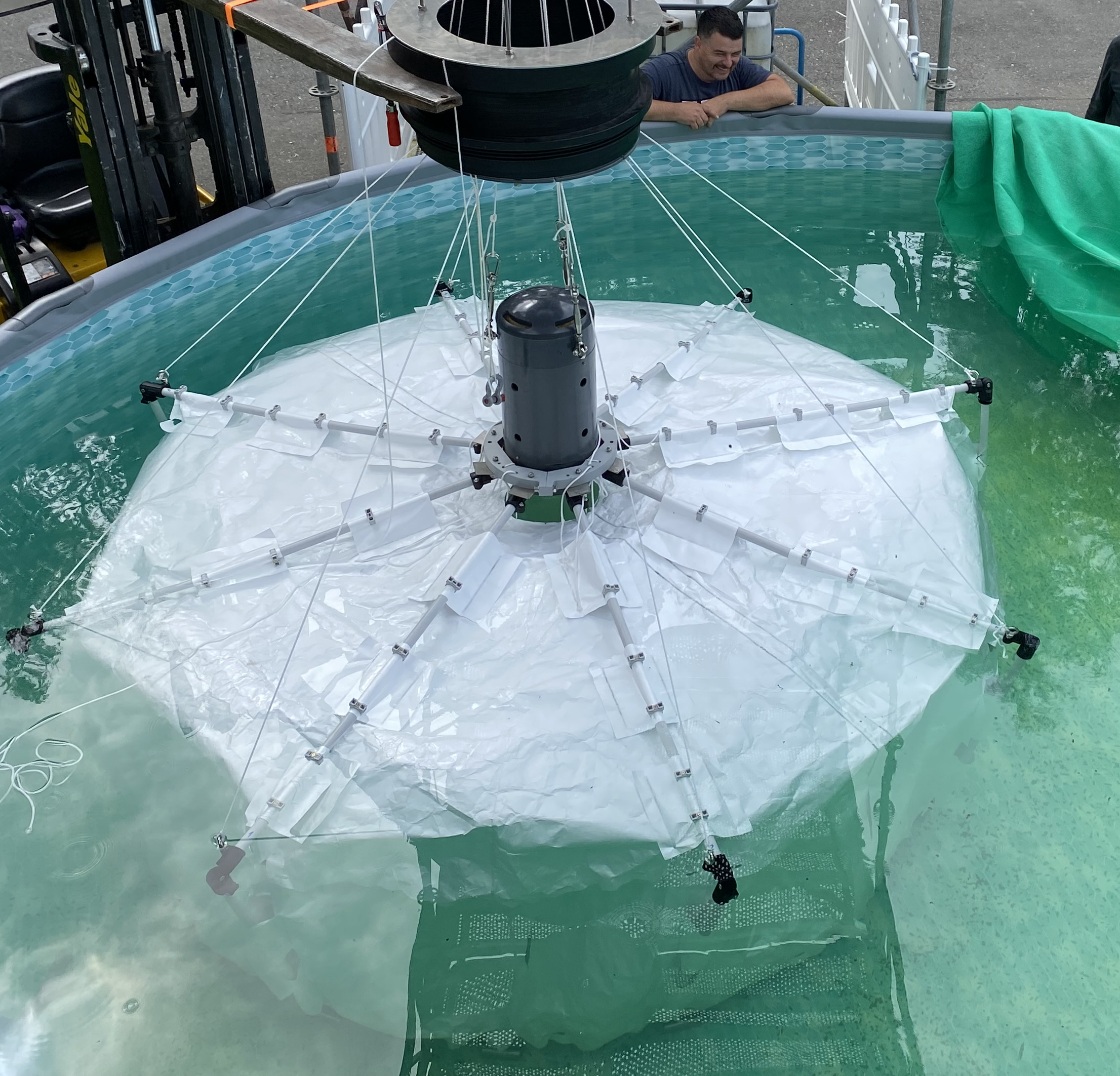} \\
    \small (a) Unfolding tests in a small pool 
  \end{tabular} \qquad
  \begin{tabular}[b]{c}
    \includegraphics[trim={0 50 0 50 cm},clip,width=.30\linewidth]{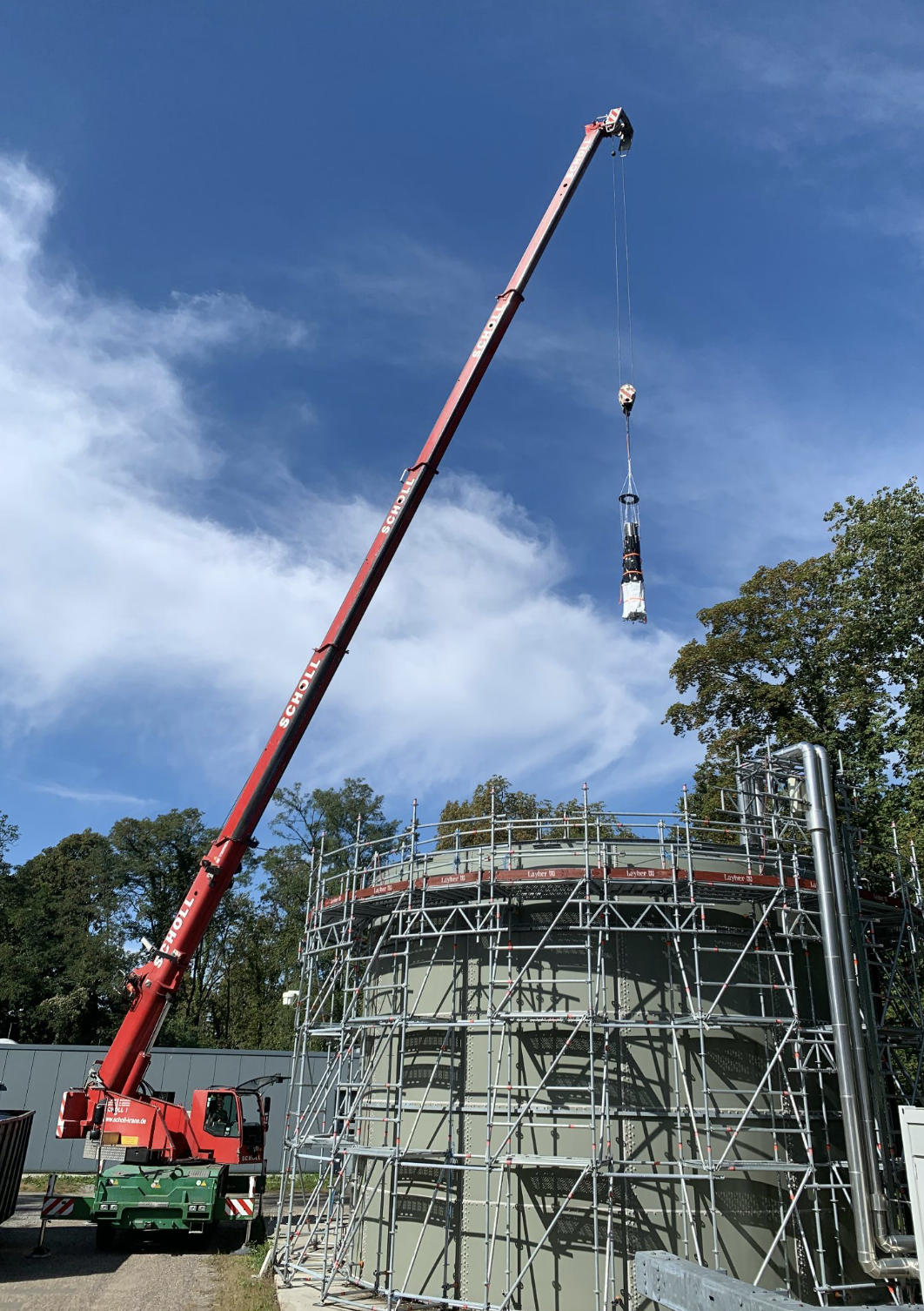} \\
    \small (b) The crane used during deployment
  \end{tabular} \\
  \begin{tabular}[b]{c}
    \includegraphics[trim={0 220 0 220 cm},clip, width=.30\linewidth]{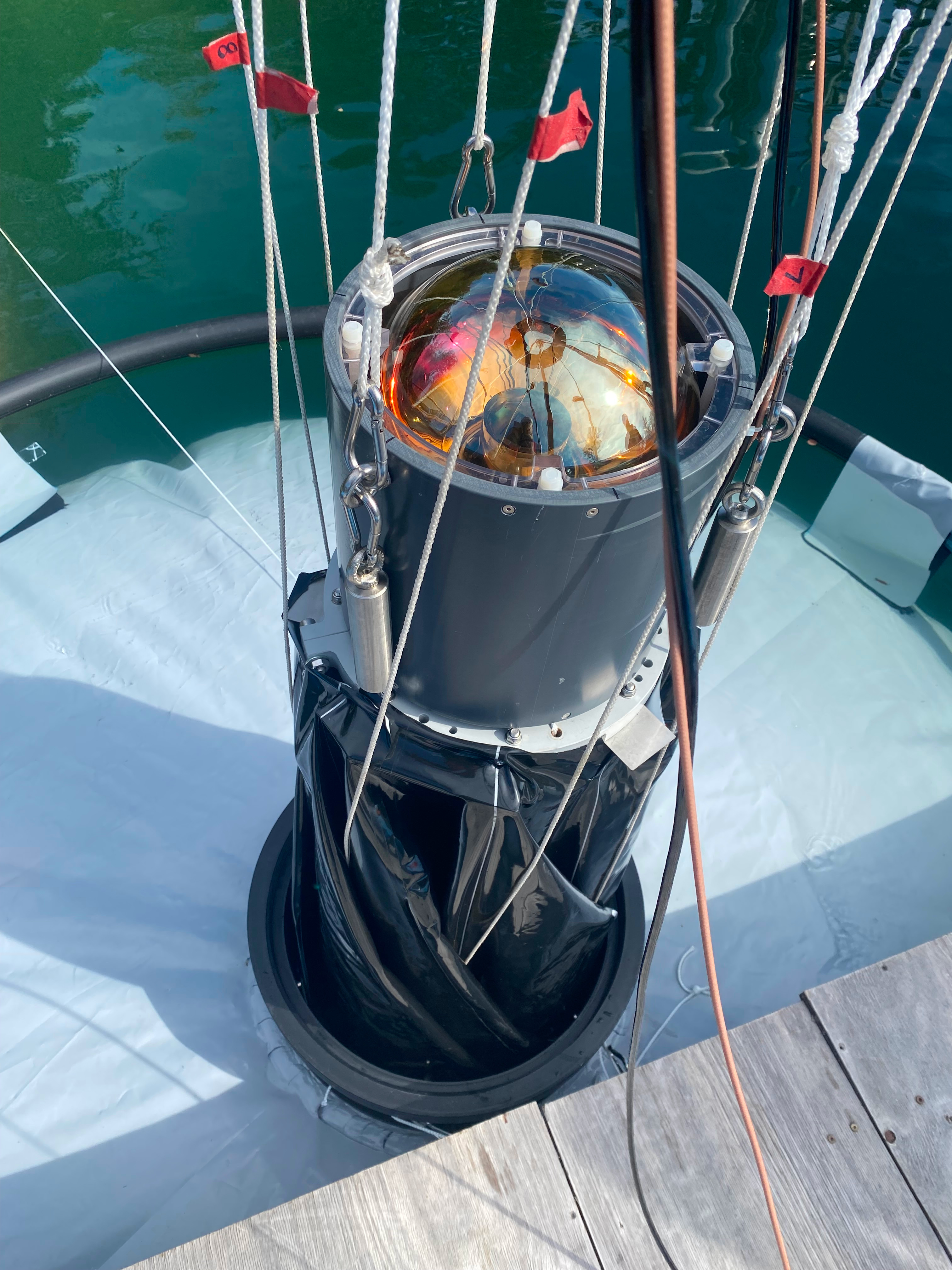} \\
    \small (c) The double-PMT structure 
  \end{tabular} \qquad
  \begin{tabular}[b]{c}
    \includegraphics[width=0.45\linewidth]{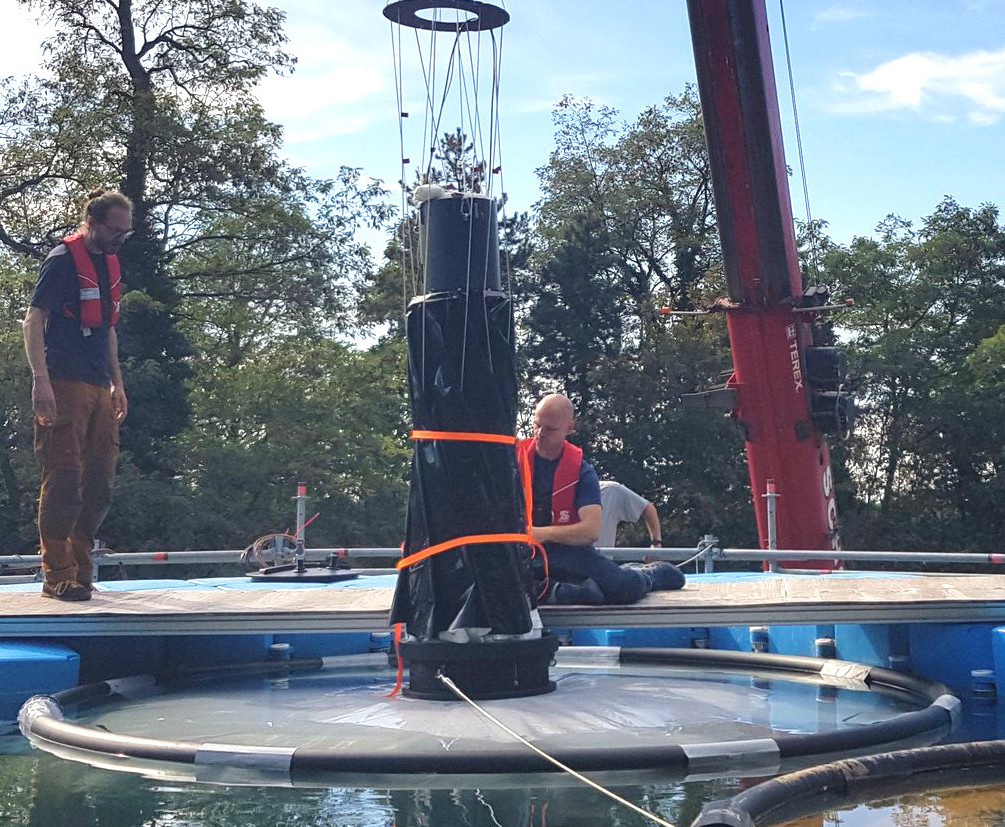} \\
    \small (d) Insertion into the pre-deployed Aquamate bladder
  \end{tabular} \\
  \begin{tabular}[b]{c}
    \includegraphics[width=.65\linewidth]{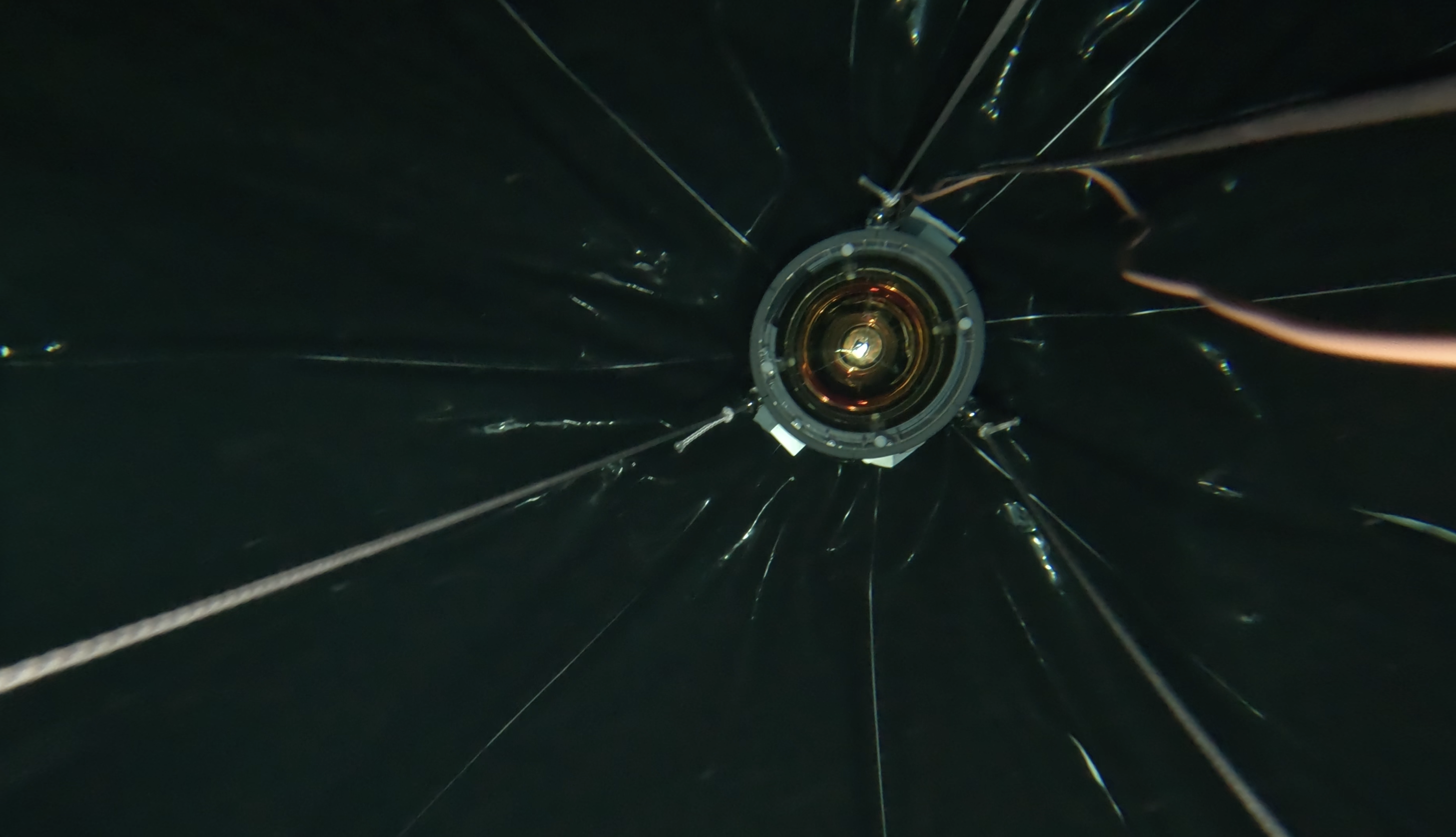} \\
    \small (e) Underwater photograph of the opened Tyvek bladder
  \end{tabular}

  \caption{Test and deployment of the Tyvek bladder. A black foil was added covering the top of the Tyvek bladder for optical isolation. The Tyvek bladder  is collapsed into a packet with diameter just below the hatch opening, is lifted with a crane to the top of the water tank, and lowered into the outer bladder. The ropes attached to the spokes cause the Tyvek bladder to unfold once fully inside the bladder. The bottom image shows an underwater photograph of the top of the opened Tyvek bladder inside the outer bladder, with the top PMT looking up to view the upper chamber.}
  \label{fig:matr_deployment}
\end{figure*}

As a first step, a single-chamber bladder (3.6\,m $\times$ 3.6\,m) from Aquamate was deployed in the test tank. The inner Tyvek~1082D volume was constructed with a foldable design that enables it to be inserted into the larger bladder, via the 60\,cm hatch. 
The Tyvek-only bladder is an octagonal cage that is connected to a double-jointed PVC-pipe umbrella that opens once the bladder is inserted into the Aquamate bladder. Fig.~\ref{fig:matr_deployment} shows the Tyvek chamber before deployment in the outer bladder. Tests for opening and closing the chamber were made first in air, and then in water as shown. The joints of the umbrella mechanism are round and there are no sharp edges to ensure there will be no tears. Furthermore, the Tyvek-only bladder is made to be smaller than the larger bladder (3). The double-PMT support (see Fig.~\ref{fig:doublePMT}) was attached to the Tyvek chamber. Then, the Tyvek chamber, along with the two PMTs, was deployed using a crane as shown in Fig.~\ref{fig:matr_deployment}.

In addition to the double-layer bladder, two instrumented black barrels with a diameter of 41\,cm and height of 75\,cm were deployed to function as muon taggers, such that the double-layer detector unit is between them (see Fig.~\ref{fig:simsetup}). Both of the barrels have an inner lining of Tyvek~1082D and are equipped with a Hamamatsu R5912-100 PMT. The bottom barrel lying at the bottom of the test pool is filled with water and functions as a WCD, meanwhile the top barrel has two layers of 4\,cm scintillators inside as the active material, instead of water (to reduce weight compared to a WCD and allow easier movement).


\subsection{Data Taking with the DLWCD prototype}

Triggering on the coincidence between the two muon taggers selects penetrating muons, and defines the path of these particles. Signals were recorded using four channels of the FlashCam readout system~\cite{WERNER201731},
for the two taggers and the top and bottom chambers. The readout system continuously records full signal waveforms with a sampling frequency of 250\,MHz, or 4\,ns per sampling step. Trigger conditions, namely a threshold of $\sim1$ photo-electron and a coincidence window of 80\,ns, are imposed in software, using the digitized signals. 

Two geometries were used for data taking, one where the top tagger was placed near to the center of the bladder, and another one where the top tagger was placed close to the edge of the bladder. When the top tagger is near to the bladder center, the path of the particles passing through is close to the double PMT unit. The bottom tagger was $\sim5.5$\,m below the water surface, while the top tagger was $\sim0.5$\,m above the water surface.

\begin{figure*}
    \centering
    \includegraphics[width=\linewidth]{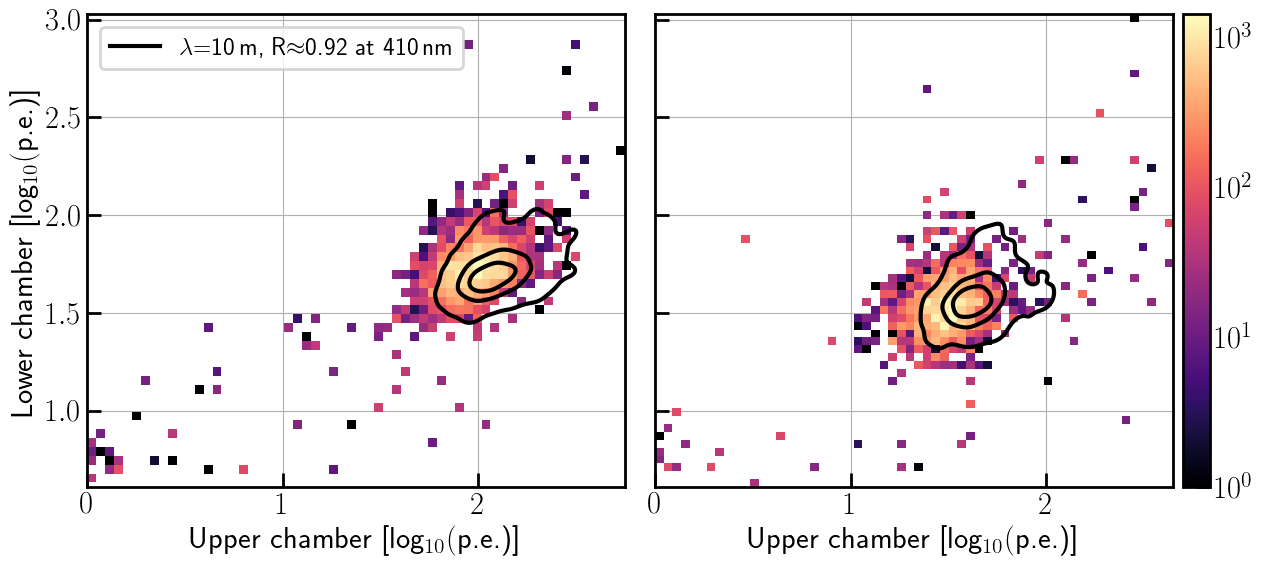}
    \caption{The measured signal distribution (histogram) compared with simulations made with HAWCSim (contour lines). \textit{Left:} The top tagger is placed close to the center of the bladder. \textit{Right:} The top tagger is moved closer to the edge of the bladder. The simulations shown are for assumed water absorption length of 10\,m at 410\,nm, and Tyvek reflectivity of R = 0.92.}
    \label{fig:data_sim}
\end{figure*}

The trigger rates were 25~mHz for the center run and 18~mHz for the edge run. The trigger rate of the top muon tagger is $\approx 410$\,Hz and the bottom muon tagger is $\approx52$\,Hz, giving an expected accidental coincidence rate of $\approx3$\,mHz. Posterior to data taking, quality cuts that require large amplitude signals from the top and bottom taggers were made to ensure that any accidental coincidence triggering is minimized. The distributions of top and bottom chamber signal amplitudes are shown in Fig.~\ref{fig:data_sim}.



The signal in the top chamber decreased significantly once the top tagger was moved to the edge of the bladder; a smaller decrease is also seen for the bottom chamber. 

In the top chamber, with its non-reflective walls, the (upward-facing) PMT basically views the direct Cherenkov light from the particle passing through the chamber; from simple geometry the amount of light hitting the PMT decreases with increasing distance of the particle track from the PMT.

For the lower chamber -- made of reflective Tyvek -- the light reaches the (downward-facing) PMT only after one or more reflections. Under ideal conditions, the light yield should be almost independent of the impact point of the track. The finite reflectivity of the Tyvek, gaps in Tyvek coverage and the finite absorption length of water will, however, result in some position dependence.


\begin{figure*}
    \centering
    \includegraphics[width=\linewidth]{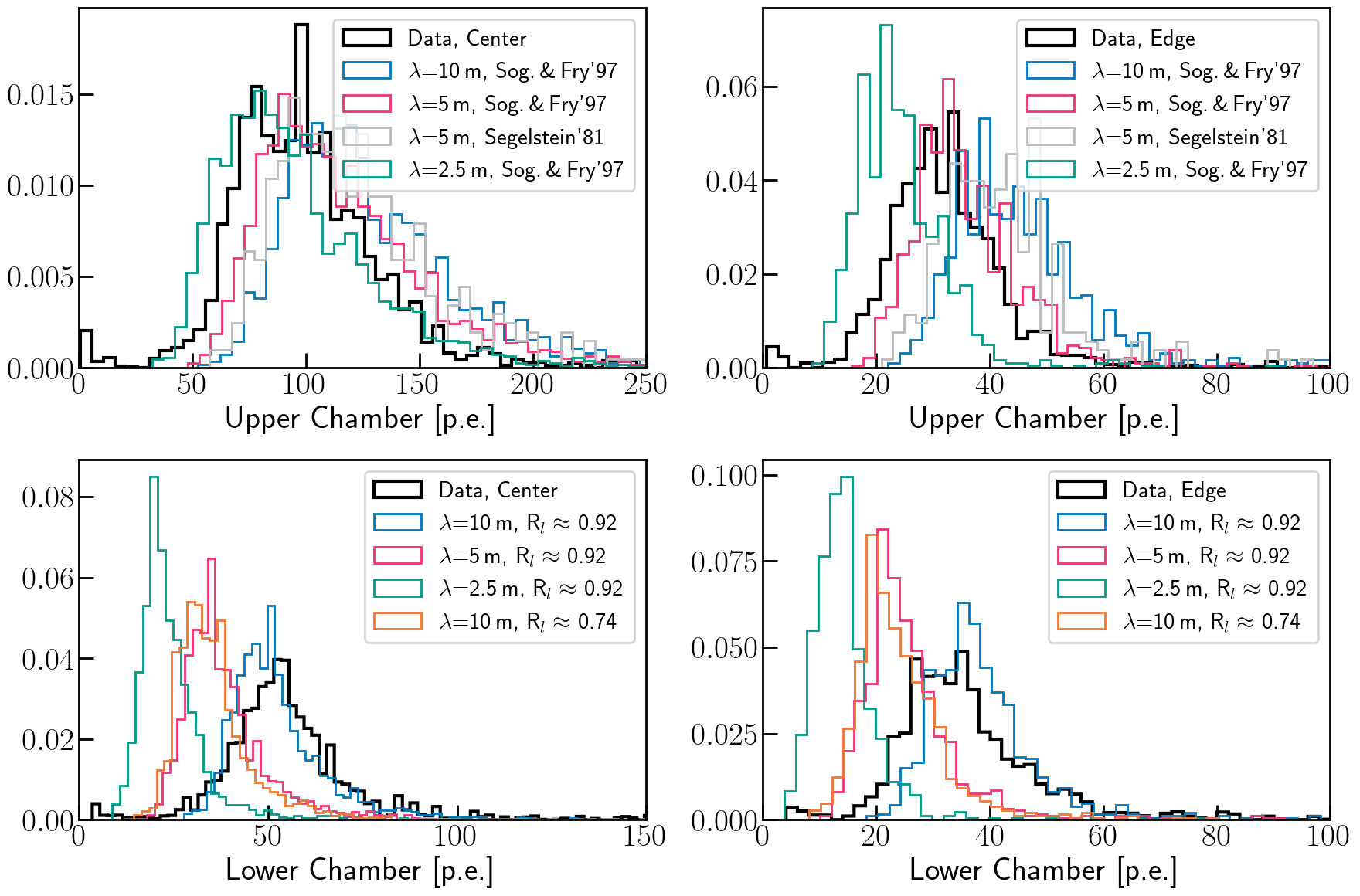}
    \caption{The measured signal distribution (black) compared with selected simulations made with HAWCSim. \textit{Left:} The top tagger is placed close to the center of the bladder. \textit{Right:} The top tagger is moved closer to the edge of the bladder. The simulations are shown for three different combinations of assumed water absorption length at 410\,nm, and Tyvek reflectivity.}
    \label{fig:data_sim_lin}
\end{figure*}
\subsection{Simulation of the Experimental Setup}

The two detector arrangements were simulated using HAWCSim. Fig.~\ref{fig:simsetup} shows the setup used in HAWCSim, where a muon traversing the test tank is shown in green, with tracing of the Cherenkov photons produced in the double-layer WCD under test. The dimensions of the double-layer bladder and two muon taggers are as measured from the real counterparts. The PMTs used in the simulations are 8-inch R5912 for the lower chamber and 10-inch R7081 for the upper chamber, following the setup with the double photo-multiplier assembly (Fig.~\ref{fig:doublePMT}). Other inputs to the simulation include the reflectivity of the inner walls, which was separately measured using an integrating sphere for the inner linings of the upper and lower chambers as mentioned in Section~\ref{sec:materials}; and the water absorption length, monitored using a Sea-Bird C-Star transmissometer at the wavelength of 410\,nm. Measurements from Pope \& Fry~\cite{popefry_water} and Sogandares \& Fry \cite{sogfry_water} are used to extrapolate the transmissometer measurement at 410~nm to cover the full spectral sensitivity of the PMTs. Several different measurements of water absorption length over a range of wavelengths exist, and although the measurements from Segelstein~\cite{Segelstein81} (1981) are very comprehensive, for these simulations concerning comparison with measured data, we use the more recent measurements of Pope \& Fry and Sogandares \& Fry (1997). The input particles used throughout the simulations were generated using the PARMA model that is implemented in the open-access software EXPACS~\cite{Sato_2016}. This model estimates terrestrial cosmic ray fluxes at a given time and place on the Earth's atmosphere, and thus provides exposure times for each input particle.

Several parameters were varied in the simulations according to their uncertainties, notably the position of the bottom tagger, the reflectivity of the inner linings, and the water absorption length. The simulations are used to estimate the expected coincidence rate and to predict the time and amplitude response of the WCD. The expected coincidence rate is found to be $\sim20$\,mHz, consistent with the data. Fig.~\ref{fig:data_sim_lin} shows measured and predicted distribution of signals in the upper and lower chambers for some of the simulations, varying water absorption length and Tyvek reflectivity.



\begin{figure*}[ht]
    \centering
    \includegraphics[width=\linewidth]{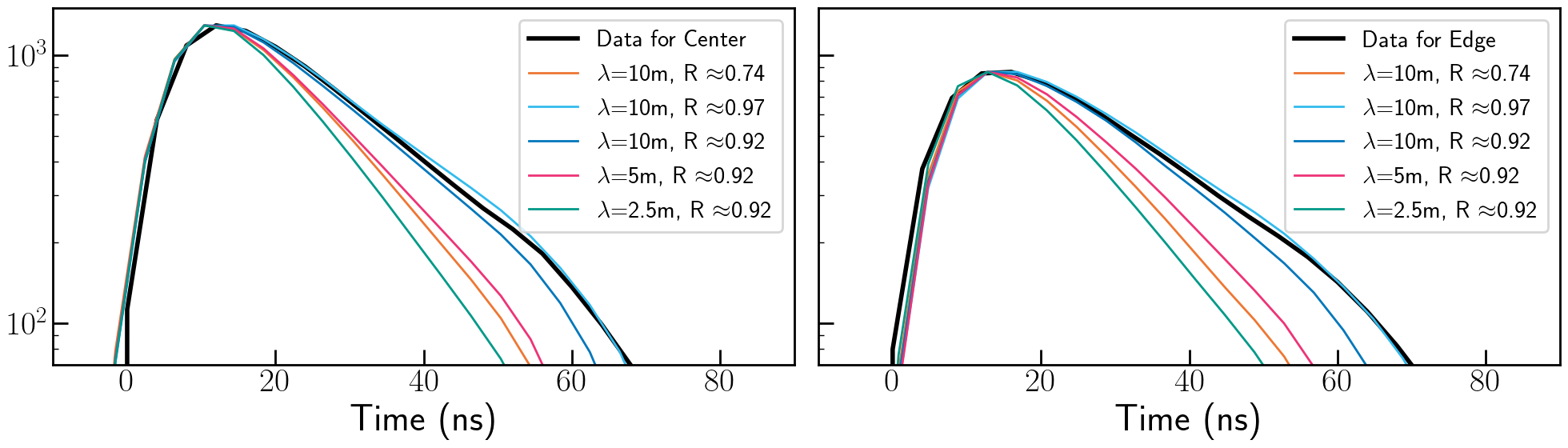}
    \caption{ Average pulse shape of the lower chamber, compared to simulations with different combinations of wall reflectivity and water absorption length.}
    \label{fig:avg_pulse}
\end{figure*}

All simulations shown are within the expected systematic uncertainties of the signal chain calibration and the PMT photon detection efficiency. 

When the water absorption length is lower, the signal in the upper and lower chambers alike gets lower, however the difference is larger for the lower chamber. Moreover, as expected, a less reflective lower chamber inner liner results in less signal captured in the lower chamber. For water absorption length spectra, within our wavelength range, simulations assuming Segelstein~\cite{Segelstein81} appear to require half the wavelength measurement at $410$\,nm to match the simulations assuming the more recent Pope \& Fry and Sogandares \& Fry spectra, one of which is shown in grey. 

While the simulations reproduce the trends seen in the data, none of the parameter combinations fully reproduces the measured distributions, possibly because of the simplified chamber geometry used in the simulations and the uncertainty coming from the PMT photon detection efficiency. 

The pulse shape of signals from the lower chamber -- recorded with 4\,ns resolution -- provides additional information and handles on wall reflectivity and water transmission. The average pulse shape of the lower chamber is shown in Fig.~\ref{fig:avg_pulse}, for data and simulations. The simulations confirm that increased Tyvek wall reflectivity or increased water transmission result in longer signals, with more late (multi-bounce) photons arriving at the PMT.

Simulations with good water transmission -- around $10$\,m at 410\,nm -- and high Tyvek reflectivity, such as shown in Fig.~\ref{fig:data_sim}, are closest to data. Overall, within uncertainties we manage to describe and characterize our measurements with the simulations.



\section{Conclusion}

The concept of a lake-based water Cherenkov detector (WCD) array for VHE gamma ray astronomy is described, discussing the advantages and challenges of the approach, the deployment and operation of WCD units, and the prototyping of a first dual-chamber detector unit.

Compared to land-based individual water tanks, advantages of the lake deployment  include cost savings, flexibility in rearranging the array configuration, and improved muon identification since the muon chamber is shielded against direct sideways entry of shower particles. Advantages are balanced by significant challenges, mostly related to demonstrating that units deployed in a lake achieve a sufficiently large lifetime (i.e. a Mean time between failures (MTBF) of more than a decade), but also the position monitoring and time calibration is challenging for WCD units that may drift under water motion, and suffer deformations.

A first prototype of a dual chamber WCD was deployed in a test tank and successfully operated. The nested bladder, or `matryoshka' approach, provides a practical solution for achieving a two-layer bladder configuration, particularly when direct production of a two-layer bladder presents challenges. The experimental data and simulations show a reasonable level of consistency in both time distribution and signal amplitudes. 


This concept of water-based WCD units also applies to artificial lakes and for ponds, which can be built on any (locally) flat surface. A pond clearly has a additional construction cost, but depending on the local conditions may still be cost-effective compared to a large number of individual tanks. The pond approach avoids problems with waves and simplifies access, but also adds new failure modes, such as a significant water leak in the pond, and requires plentiful water availability (also to balance evaporation). In context of SWGO, the possibility of a lake deployment of detector units is considered for a future extension of the array, after the completion of the main array which will be land based. 

Future steps to validate the concept of lake-based WCD arrays include a demonstration of unit detector behavior under wave motion, and the deployment of an array with multiple unit detectors under realistic conditions.



\section*{Acknowledgements}

We would like to thank our colleagues in the SWGO collaboration for valuable discussions and the insightful questions that they posed throughout this work. Special thanks go to Dr Samridha Kunwar for his contributions and support with the simulations, and the technical team at MPIK for their invaluable assistance with the prototyping.


\vspace{1cm}

\bibliographystyle{elsarticle-num-names}
\bibliography{lake_bibs}

\begin{thebibliography}{28}
\expandafter\ifx\csname natexlab\endcsname\relax\def\natexlab#1{#1}\fi
\providecommand{\url}[1]{\texttt{#1}}
\providecommand{\href}[2]{#2}
\providecommand{\path}[1]{#1}
\providecommand{\DOIprefix}{doi:}
\providecommand{\ArXivprefix}{arXiv:}
\providecommand{\URLprefix}{URL: }
\providecommand{\Pubmedprefix}{pmid:}
\providecommand{\doi}[1]{\href{http://dx.doi.org/#1}{\path{#1}}}
\providecommand{\Pubmed}[1]{\href{pmid:#1}{\path{#1}}}
\providecommand{\bibinfo}[2]{#2}
\ifx\xfnm\relax \def\xfnm[#1]{\unskip,\space#1}\fi
\bibitem[{Abeysekara et~al.(2017)}]{Abeysekara_2017}
\bibinfo{author}{A.~U. Abeysekara}, et~al.,
\newblock \bibinfo{title}{Observation of the crab nebula with the hawc
  gamma-ray observatory},
\newblock \bibinfo{journal}{The Astrophysical Journal} \bibinfo{volume}{843}
  (\bibinfo{year}{2017}) \bibinfo{pages}{39}. \URLprefix
  \url{https://doi.org/10.3847/1538-4357/aa7555}.
  \DOIprefix\doi{10.3847/1538-4357/aa7555}.
\bibitem[{{Ma} and et~al(2022)}]{2022ChPhC..46c0001M}
\bibinfo{author}{X.-H. {Ma}}, \bibinfo{author}{et~al},
\newblock \bibinfo{title}{{LHAASO Instruments and Detector technology}},
\newblock \bibinfo{journal}{Chinese Physics C} \bibinfo{volume}{46}
  (\bibinfo{year}{2022}) \bibinfo{pages}{030001}.
  \DOIprefix\doi{10.1088/1674-1137/ac3fa6}.
\bibitem[{{Belolaptikov} et~al.(1997)}]{1997APh.....7..263B}
\bibinfo{author}{I.~A. {Belolaptikov}}, et~al.,
\newblock \bibinfo{title}{{The Baikal underwater neutrino telescope: Design,
  performance, and first results}},
\newblock \bibinfo{journal}{Astroparticle Physics} \bibinfo{volume}{7}
  (\bibinfo{year}{1997}) \bibinfo{pages}{263--282}.
  \DOIprefix\doi{10.1016/S0927-6505(97)00022-4}.
\bibitem[{{Kaneko} and et~al(1983)}]{1983ICRC...11..428K}
\bibinfo{author}{T.~{Kaneko}}, \bibinfo{author}{et~al},
\newblock \bibinfo{title}{{Acoustic and Vlf-Elf Radio Detections of Super Giant
  Air Showers}},
\newblock in: \bibinfo{booktitle}{International Cosmic Ray Conference},
  volume~\bibinfo{volume}{11} of \textit{\bibinfo{series}{International Cosmic
  Ray Conference}}, \bibinfo{year}{1983}, p. \bibinfo{pages}{428}.
\bibitem[{Hinton(2021)}]{Hinton_2021}
\bibinfo{author}{J.~Hinton},
\newblock \bibinfo{title}{The southern wide-field gamma-ray observatory: Status
  and prospects},
\newblock \bibinfo{journal}{Proceedings of 37th International Cosmic Ray
  Conference PoS(ICRC2021)}  (\bibinfo{year}{2021}).
\bibitem[{Goksu and Hofmann(2021)}]{Goksu:2021Tw}
\bibinfo{author}{H.~Goksu}, \bibinfo{author}{W.~Hofmann},
\newblock \bibinfo{title}{{Lake Deployment of Southern Wide-field Gamma-ray
  Observatory (SWGO) Detector Units}},
\newblock \bibinfo{journal}{PoS} \bibinfo{volume}{ICRC2021}
  (\bibinfo{year}{2021}) \bibinfo{pages}{708}.
  \DOIprefix\doi{10.22323/1.395.0708}.
\bibitem[{Goksu(2023)}]{Goksu:2023Pk}
\bibinfo{author}{H.~Goksu},
\newblock \bibinfo{title}{{Updates on Lake Deployment of Southern Wide-field
  Gamma-ray Observatory (SWGO) Detector Units}},
\newblock \bibinfo{journal}{PoS} \bibinfo{volume}{ICRC2023}
  (\bibinfo{year}{2023}) \bibinfo{pages}{653}.
  \DOIprefix\doi{10.22323/1.444.0653}.
\bibitem[{{Halzen} and {Stanev}(1995)}]{1995hep.ph....7362H}
\bibinfo{author}{F.~{Halzen}}, \bibinfo{author}{T.~{Stanev}},
\newblock \bibinfo{title}{{Gamma Ray Astronomy with Underground Detectors}},
\newblock \bibinfo{journal}{arXiv e-prints}  (\bibinfo{year}{1995})
  \bibinfo{pages}{hep--ph/9507362}.
  \DOIprefix\doi{10.48550/arXiv.hep-ph/9507362}.
  \href{http://arxiv.org/abs/hep-ph/9507362}{{\tt arXiv:hep-ph/9507362}}.
\bibitem[{{Ahlen} and {MACRO Collaboration}(1993)}]{1993ApJ...412..301A}
\bibinfo{author}{S.~{Ahlen}}, \bibinfo{author}{{MACRO Collaboration}},
\newblock \bibinfo{title}{{Muon Astronomy with the MACRO Detector}},
\newblock \bibinfo{journal}{The Astrophysical Journal} \bibinfo{volume}{412}
  (\bibinfo{year}{1993}) \bibinfo{pages}{301}. \DOIprefix\doi{10.1086/172921}.
\bibitem[{{SWGO Collaboration} et~al.(2022){SWGO Collaboration}, {Doro}, and
  et~al.}]{2022icrc.confE.689D}
\bibinfo{author}{{SWGO Collaboration}}, \bibinfo{author}{M.~{Doro}},
  \bibinfo{author}{et~al.},
\newblock \bibinfo{title}{{The search for high altitude sites in South America
  for the SWGO detector}},
\newblock in: \bibinfo{booktitle}{37th International Cosmic Ray Conference.
  12-23 July 2021. Berlin}, \bibinfo{year}{2022}, p. \bibinfo{pages}{689}.
\bibitem[{Abreu and et~al.(2023)}]{Abreu:2023WI}
\bibinfo{author}{P.~Abreu}, \bibinfo{author}{et~al.},
\newblock \bibinfo{title}{{An update on site search activities for SWGO}},
\newblock in: \bibinfo{booktitle}{Proceedings of 38th International Cosmic Ray
  Conference {\textemdash} PoS(ICRC2023)}, volume \bibinfo{volume}{444},
  \bibinfo{year}{2023}, p. \bibinfo{pages}{752}.
  \DOIprefix\doi{10.22323/1.444.0752}.
\bibitem[{{Sobel} et~al.(1990){Sobel}, {Adams}, {Bond}, {Bratton}, {Burnett},
  {Chaloupka}, {Cherry}, {Coleman}, {Ellison}, {Gaidos}, {Goodman}, {Gurr},
  {Guzik}, {Haines}, {Kielczewska}, {Kropp}, {Lane}, {Lieber}, {Loeffler},
  {Nagle}, {Nelson}, {Potter}, {Price}, {Reines}, {Rollefson}, {Schultz},
  {Sembroski}, {Sobel}, {Steinberg}, {Svoboda}, {Tripp}, {Wefel}, {Wilkes},
  {Wilson}, {Wold}, {Yodh}, and {GRANDE Collaboration}}]{1990NuPhS..14..125S}
\bibinfo{author}{H.~{Sobel}}, \bibinfo{author}{A.~{Adams}},
  \bibinfo{author}{R.~{Bond}}, \bibinfo{author}{C.~B. {Bratton}},
  \bibinfo{author}{T.~{Burnett}}, \bibinfo{author}{V.~{Chaloupka}},
  \bibinfo{author}{M.~{Cherry}}, \bibinfo{author}{L.~{Coleman}},
  \bibinfo{author}{S.~B. {Ellison}}, \bibinfo{author}{J.~{Gaidos}},
  \bibinfo{author}{J.~{Goodman}}, \bibinfo{author}{H.~{Gurr}},
  \bibinfo{author}{T.~G. {Guzik}}, \bibinfo{author}{T.~J. {Haines}},
  \bibinfo{author}{D.~{Kielczewska}}, \bibinfo{author}{W.~{Kropp}},
  \bibinfo{author}{C.~{Lane}}, \bibinfo{author}{M.~{Lieber}},
  \bibinfo{author}{F.~{Loeffler}}, \bibinfo{author}{D.~{Nagle}},
  \bibinfo{author}{M.~{Nelson}}, \bibinfo{author}{M.~{Potter}},
  \bibinfo{author}{L.~R. {Price}}, \bibinfo{author}{F.~{Reines}},
  \bibinfo{author}{A.~{Rollefson}}, \bibinfo{author}{J.~{Schultz}},
  \bibinfo{author}{G.~{Sembroski}}, \bibinfo{author}{H.~{Sobel}},
  \bibinfo{author}{R.~{Steinberg}}, \bibinfo{author}{R.~{Svoboda}},
  \bibinfo{author}{R.~{Tripp}}, \bibinfo{author}{J.~{Wefel}},
  \bibinfo{author}{R.~J. {Wilkes}}, \bibinfo{author}{C.~{Wilson}},
  \bibinfo{author}{D.~{Wold}}, \bibinfo{author}{G.~{Yodh}},
  \bibinfo{author}{{GRANDE Collaboration}},
\newblock \bibinfo{title}{{The GRANDE detector}},
\newblock \bibinfo{journal}{Nuclear Physics B Proceedings Supplements}
  \bibinfo{volume}{14} (\bibinfo{year}{1990}) \bibinfo{pages}{125--142}.
  \DOIprefix\doi{10.1016/0920-5632(90)90409-N}.
\bibitem[{{Smith}(2005)}]{2005ICRC...10..227S}
\bibinfo{author}{A.~J. {Smith}},
\newblock \bibinfo{title}{{The MILAGRO Gamma Ray Observatory}},
\newblock in: \bibinfo{editor}{B.~S. {Acharya}}, \bibinfo{editor}{S.~{Gupta}},
  \bibinfo{editor}{P.~{Jagadeesan}}, \bibinfo{editor}{A.~{Jain}},
  \bibinfo{editor}{S.~{Karthikeyan}}, \bibinfo{editor}{S.~{Morris}},
  \bibinfo{editor}{S.~{Tonwar}} (Eds.), \bibinfo{booktitle}{29th International
  Cosmic Ray Conference (ICRC29), Volume 10}, volume~\bibinfo{volume}{10} of
  \textit{\bibinfo{series}{International Cosmic Ray Conference}},
  \bibinfo{year}{2005}, p. \bibinfo{pages}{227}.
\bibitem[{{Kunwar} and et~al(2023)}]{2023NIMPA105068138K}
\bibinfo{author}{S.~{Kunwar}}, \bibinfo{author}{et~al},
\newblock \bibinfo{title}{{A double-layered Water Cherenkov Detector array for
  Gamma-ray astronomy}},
\newblock \bibinfo{journal}{Nuclear Instruments and Methods in Physics Research
  A} \bibinfo{volume}{1050} (\bibinfo{year}{2023}) \bibinfo{pages}{168138}.
  \DOIprefix\doi{10.1016/j.nima.2023.168138}.
  \href{http://arxiv.org/abs/2209.09305}{{\tt arXiv:2209.09305}}.
\bibitem[{{Pierre Auger Collaboration} et~al.(2008){Pierre Auger
  Collaboration}, {Allekotte}, and et~al}]{2008NIMPA.586..409P}
\bibinfo{author}{{Pierre Auger Collaboration}},
  \bibinfo{author}{I.~{Allekotte}}, \bibinfo{author}{et~al},
\newblock \bibinfo{title}{{The surface detector system of the Pierre Auger
  Observatory}},
\newblock \bibinfo{journal}{Nuclear Instruments and Methods in Physics Research
  A} \bibinfo{volume}{586} (\bibinfo{year}{2008}) \bibinfo{pages}{409--420}.
  \DOIprefix\doi{10.1016/j.nima.2007.12.016}.
  \href{http://arxiv.org/abs/0712.2832}{{\tt arXiv:0712.2832}}.
\bibitem[{{Abeysekara}(2023)}]{2023NIMPA105268253A}
\bibinfo{author}{H.~{Abeysekara}, A.~U. et~al},
\newblock \bibinfo{title}{{The High-Altitude Water Cherenkov (HAWC) observatory
  in M{\'e}xico: The primary detector}},
\newblock \bibinfo{journal}{Nuclear Instruments and Methods in Physics Research
  A} \bibinfo{volume}{1052} (\bibinfo{year}{2023}) \bibinfo{pages}{168253}.
  \DOIprefix\doi{10.1016/j.nima.2023.168253}.
  \href{http://arxiv.org/abs/2304.00730}{{\tt arXiv:2304.00730}}.
\bibitem[{Wang et~al.(2020)Wang, Li, Xiao, Zhang, Feng, Wang, Li, Zuo, Cheng,
  Wu, Zhang, He, and Jia}]{WANG2020163416}
\bibinfo{author}{H.~Wang}, \bibinfo{author}{C.~Li}, \bibinfo{author}{G.~Xiao},
  \bibinfo{author}{Y.~Zhang}, \bibinfo{author}{S.~Feng},
  \bibinfo{author}{L.~Wang}, \bibinfo{author}{X.~Li}, \bibinfo{author}{X.~Zuo},
  \bibinfo{author}{N.~Cheng}, \bibinfo{author}{W.~Wu},
  \bibinfo{author}{Y.~Zhang}, \bibinfo{author}{H.~He},
  \bibinfo{author}{H.~Jia},
\newblock \bibinfo{title}{Measuring the optical parameters for lhaaso-md},
\newblock \bibinfo{journal}{Nuclear Instruments and Methods in Physics Research
  Section A: Accelerators, Spectrometers, Detectors and Associated Equipment}
  \bibinfo{volume}{956} (\bibinfo{year}{2020}) \bibinfo{pages}{163416}.
  \URLprefix
  \url{https://www.sciencedirect.com/science/article/pii/S016890022030022X}.
  \DOIprefix\doi{https://doi.org/10.1016/j.nima.2020.163416}.
\bibitem[{{Bosboom } and {Stive}(2023)}]{waves_coastal}
\bibinfo{author}{J.~{Bosboom }}, \bibinfo{author}{M.~{Stive}},
  \bibinfo{title}{Coastal Dynamics}, \bibinfo{publisher}{Delft University of
  Technology}, \bibinfo{year}{2023}.
\bibitem[{of~Engineers(2008)}]{coastal_engineering}
\bibinfo{author}{U.~A.~C. of~Engineers}, \bibinfo{title}{Coastal Engineering
  Manual, EM 1110-2-1100 Water Wave Mechanics (Part II)},
  \bibinfo{publisher}{US Army Corps of Engineers}, \bibinfo{year}{2008}.
\bibitem[{{Stewart}(2000)}]{oceanography}
\bibinfo{author}{R.~{Stewart}}, \bibinfo{title}{Introduction to Physical
  Oceanography}, \bibinfo{publisher}{Texas A \& M University},
  \bibinfo{year}{2000}.
\bibitem[{{Børkja}(2015)}]{master_norway}
\bibinfo{author}{J.~{Børkja}}, \bibinfo{title}{Dynamic analysis of floating
  dock structures}, \bibinfo{year}{2015}.
\bibitem[{Agostinelli et~al.(2003)}]{GEANT4_2002}
\bibinfo{author}{S.~Agostinelli}, et~al. (\bibinfo{collaboration}{GEANT4}),
\newblock \bibinfo{title}{{GEANT4--a simulation toolkit}},
\newblock \bibinfo{journal}{Nucl. Instrum. Meth. A} \bibinfo{volume}{506}
  (\bibinfo{year}{2003}) \bibinfo{pages}{250--303}.
  \DOIprefix\doi{10.1016/S0168-9002(03)01368-8}.
\bibitem[{Segelstein(1981)}]{Segelstein81}
\bibinfo{author}{D.~J. Segelstein}, \bibinfo{title}{The complex refractive
  index of water}, Ph.D. thesis, University of Missouri-Kansas City,
  \bibinfo{year}{1981}.
\bibitem[{Werner and Nellen(2021)}]{Werner:2021d/}
\bibinfo{author}{F.~Werner}, \bibinfo{author}{L.~Nellen},
\newblock \bibinfo{title}{{Technological options for the Southern Wide-field
  Gamma-ray Observatory (SWGO) and current design status}},
\newblock in: \bibinfo{booktitle}{Proceedings of 37th International Cosmic Ray
  Conference {\textemdash} PoS(ICRC2021)}, volume \bibinfo{volume}{395},
  \bibinfo{year}{2021}, p. \bibinfo{pages}{714}.
  \DOIprefix\doi{10.22323/1.395.0714}.
\bibitem[{Werner et~al.(2017)}]{WERNER201731}
\bibinfo{author}{F.~Werner}, et~al.,
\newblock \bibinfo{title}{Performance verification of the flashcam prototype
  camera for the cherenkov telescope array},
\newblock \bibinfo{journal}{Nuclear Instruments and Methods in Physics Research
  Section A: Accelerators, Spectrometers, Detectors and Associated Equipment}
  \bibinfo{volume}{876} (\bibinfo{year}{2017}) \bibinfo{pages}{31--34}.
  \DOIprefix\doi{https://doi.org/10.1016/j.nima.2016.12.056},
  \bibinfo{note}{the 9th international workshop on Ring Imaging Cherenkov
  Detectors (RICH2016)}.
\bibitem[{{Pope} and {Fry}(1997)}]{popefry_water}
\bibinfo{author}{R.~{Pope}}, \bibinfo{author}{E.~{Fry}},
\newblock \bibinfo{title}{Absorption spectrum (380-700 nm) of pure water. ii.
  integrating cavity measurements},
\newblock \bibinfo{journal}{Appl. Opt.,36, 8710-8723}  (\bibinfo{year}{1997}).
\bibitem[{{Sogandares} and {Fry}(1997)}]{sogfry_water}
\bibinfo{author}{F.~{Sogandares}}, \bibinfo{author}{E.~{Fry}},
\newblock \bibinfo{title}{Absorption spectrum (340- 640 nm) of pure water. i.
  photothermal measurements},
\newblock \bibinfo{journal}{Appl. Opt.,36, 8699-8709}  (\bibinfo{year}{1997}).
\bibitem[{Sato(2016)}]{Sato_2016}
\bibinfo{author}{T.~Sato},
\newblock \bibinfo{title}{Analytical model for estimating the zenith angle
  dependence of terrestrial cosmic ray fluxes},
\newblock \bibinfo{journal}{PLOS ONE} \bibinfo{volume}{11}
  (\bibinfo{year}{2016}) \bibinfo{pages}{1--22}. \URLprefix
  \url{https://doi.org/10.1371/journal.pone.0160390}.
  \DOIprefix\doi{10.1371/journal.pone.0160390}.

\end{thebibliography}

\end{document}